\documentclass[fleqn,usenatbib]{mnras}

\usepackage[T1]{fontenc}

\DeclareRobustCommand{\VAN}[3]{#2}
\let\VANthebibliography\thebibliography
\def\thebibliography{\DeclareRobustCommand{\VAN}[3]{##3}\VANthebibliography}

\usepackage{graphicx}	% Including figure files
\usepackage{amsmath}	% Advanced maths commands
\usepackage{amssymb}	% Extra maths symbols
\usepackage{upgreek}

\usepackage{color}
\usepackage{aas_macros}
\usepackage{graphicx}
\usepackage[normalem]{ulem}

\newcommand{\kms}{{\rm {km\, s^{-1}}}}
\newcommand{\msun}{{\rm M_{\sun }}}
\newcommand{\Mo}{\msun}
\newcommand{\Ro}{{\rm R_{\sun }}}

\newcommand{\Teff}{T_{\rm eff}}
\newcommand{\ud}{{\rm d}}

\def\Mc{M_{\rm c}}
\def\Rc{R_{\rm c}}
\def\r{{\tilde{r}}}
\def\vt{{\tilde{v}}}

\title[Stellar collisions in the Galactic Centre I]{Close stellar encounters at the Galactic Centre I: The effect on the observed stellar populations}
\author[A.~Mastrobuono-Battisti et al.]{Alessandra Mastrobuono-Battisti$^{1}$\thanks{E-mail:
mastrobuono@astro.lu.se}, Ross~P.~Church$^{1}$ and Melvyn~B.~Davies$^{1,2}$\\
$^{1}$Department of Astronomy and Theoretical Physics, Lund Observatory, Box 43, SE--221 00, Lund, Sweden.\\
$^{2}$Centre for Mathematical Sciences, Lund University, Box 118, SE--221 00, Lund, Sweden.\\
}
\begin{document}

\date{Accepted XXX. Received XXX; in original form XXX}

\pagerange{\pageref{firstpage}--\pageref{lastpage}} \pubyear{2021}

\maketitle

\label{firstpage}

\begin{abstract}
We model the effects of collisions and close encounters on the stellar populations observed in the Milky Way nuclear stellar cluster (NSC).  Our analysis is based on $N$-body simulations in which the NSC forms by accretion of massive stellar clusters around a supermassive black hole.  We attach stellar populations to our $N$-body particles and follow the evolution of their stars, and the rate of collisions and close encounters.  The most common encounters are collisions between pairs of main-sequence stars, which lead to mergers: destructive collisions between main-sequence stars and compact objects are rare.  We find that the effects of collisions on the stellar populations are small for three reasons.  First, our models possess a core which limits the maximum stellar density.  Secondly, the velocity dispersion in the NSC is similar to the surface escape velocities of the stars, which minimises the collision rate.  Finally, whilst collisions between main-sequence stars destroy bright giants by accelerating their evolution, they also create them by accelerating the evolution of lower-mass stars. These two effects approximately cancel out.  We also investigate whether the G2 cloud could be a fuzzball: a compact stellar core which has accreted a tenuous envelope in a close encounter with a red giant.  We conclude that fuzzballs with cores below $2\,\msun$ have thermal times-scales too short to reproduce G2. A fuzzball with a black-hole core could reproduce the surface properties of G2 but the production rate of such objects in our model is low.
\end{abstract}


\section{Introduction}

Nuclear star clusters (NSCs) are among the densest stellar systems in the Universe \citep[see][for a recent and comprehensive review of the properties and evolution of NSCs]{Neumayer20}. With half-light radii of a few parsecs and masses of $10^6$-$10^7\,\msun$, they are located at the dynamical centres of the majority of observed galaxies \citep{Boeker04, Cote06, Boeker10, Neumayer11, Turner12, Georgiev14, denbrok14, Neumayer20}. Many of these NSCs coexist with a central supermassive black hole \citep[SMBH, see e.g.][]{Neumayer12, Nguyen19}. This is the case for our Galaxy, whose NSC hosts at its centre SgrA*, an SMBH of $4.3\times10^6~M_\odot$ \citep{GH98, EI05, GI09, BGS16, GP17}. The Galactic NSC has a total mass of $\sim 2.5\times10^7~M_\odot$ and a projected half-light radius of $4.2$~pc \citep{Schoedel14a, Schoedel14b}. 
Thanks to multi-epoch high-quality adaptive optics observations, \cite{Schoedel20} found that 80\% of the total stellar mass of the Galactic NSC formed 10\,Gyr ago. After a quiescent phase that lasted around 5\,Gyr, there has been another star formation episode that generated 15\% of the observed stellar mass. Finally, a few percent of the mass of the NSC formed 100\,Myr ago.  

Although the NSC properties and evolutionary history seem to be correlated to those of their host galaxies, there is not yet a clear consensus on their formation mechanism. 
One hypothesis is that NSCs formed in situ, from gas that fragmented after flowing into the centres of the galaxies \citep{Loose82, Milosavljevic04, Schinnerer06, Schinnerer08}. This scenario has been tested using hydrodynamical and $N$-body simulations \citep[see e.g.][]{LE03, Nayakshin05, Paumard06, Hobbs09, Mapelli12, Mastrobuono19} that mostly focused on reproducing the disc of young stars observed within the central 0.5~pc of the Galactic NSC \cite[see e.g.][]{Genzel10}.

Another alternative mechanism, proposed by \cite{Tremaine75}, involves the merger of massive, dense stellar clusters, similar to globular clusters, that spiralled into the centre of the galaxy because of the action of dynamical friction  \citep[see also][]{CD93}. This mechanism has been studied in detail using $N$-body simulations \citep[see e.g.][]{Antonini12, Gnedin14, Arcasedda15, AN15}. Some of the simulations were tailored to the Milky Way and could produce an NSC structurally and dynamically compatible with that observed at the Galactic centre \citep{Antonini12, Mastrobuono14, PMB14, Tsatsi17, Abbate18}. The two proposed scenarios could both partake in the formation of NSCs. It has also been suggested that clusters could be the potential source of gas from which new stars form in galactic nuclei \citep{Guillard16}. 

Using recent observations of the Galactic centre, \cite{Schoedel20} seem to constrain the contribution of classical globular clusters to a small fraction of the total mass. Recently, \cite{Feldmeier14}, \cite{Feldmeier17} and \cite{Do20} found a low metallicity substructure with kinematical properties that seem to suggest that it could have been generated by a recent stellar cluster merger at the Galactic centre \citep{Tsatsi17, Arcasedda20}. However, a clear consensus on the origin of NSCs is still missing.

%%Summary of observations of giant depletion at the Galactic Centre (adding some possible theoretical explanations)
 
Following their dynamical evolution, stars around an SMBH should relax and reach a cusp-like distribution in density space \citep{BW76}. However, observations have shown that red giant (RG) stars show a core-like distribution in the inner 0.5~pc (12'') of the Galaxy, with a flat or even radially decreasing projected profile \citep[see e.g.][]{Sellgren90, GT96, BS09, DG09, YZ12}. Several dynamical processes have been suggested to explain this so called ``missing RG'' problem. For example, \cite{GT96} suggested that the core could be the result of  collisions between  RGs and main sequence (MS) stars. These kind of collisions should be frequent in the extremely dense central regions of the NSC. This scenario has been explored theoretically \citep{Davies98,Alexander99,BaileyDavies99,Dale+09}; these studies found that collisions would produce a NSC core too small with respect to the observations. 
%%These reference need to be extended since they are impotant for this work
Other mechanisms invoked to explain the discrepancy between models and observations include the action of an intermediate mass black hole or the infall of a star cluster \citep[see e.g.][]{Gualandris12, Antonini12}. These scenarios imply that the Galaxy and the NSC went through processes not supported by other observables (e.g. a major merger or frequent and recent stellar cluster infalls). More recent works \citep{GaSc18, Schoedel18, Habibi19, Schoedel20}, using improved observational techniques, revised our knowledge of the stellar distribution in the central region of the NSC, finding a steeper density profile than seen in previous studies. In particular, \cite{GaSc18} showed that fainter late-type stars have a spatial distribution well described by a power-law with index $\simeq-1.43$. However, brighter RG stars show a shallower cusp, with power-law index $\simeq -1.2$. This suggests that $\sim 100$ RGs could be missing within 0.3~pc from Sgr A*. Similarly, \cite{Habibi19}  found a core-like structure only for the brightest RG stars. However, the number of missing RGs in their case seems to be smaller than 100.
Based on results by \cite{AS14}, \cite{AS20} illustrated how the interaction between the stars in the cusp and the clumpy gaseous discs that generated new stellar populations at the Galactic centre could have stripped the envelope of the RGs making them invisible. \cite{ZA20} recently suggested that the lack of RGs could be the result of the interaction between these stars and the nuclear jet that left the observed $\gamma$-ray Fermi bubbles as an imprint of its past action.

Another open question is the nature of G2, a cloud-like object orbiting the Galactic Centre. This cloud has been observed several times in various wavelengths \citep[e.g.][]{Gillessen12,Phifer13,Hora14,Witzel+14,Bower15,Pfuhl15,Tsuboi15, Valencia15, Plewa17}.
There are two main classes of proposed origins for this object. In the former G2 is considered as purely a gas cloud. In the latter, G2 is the result of the ejection of gas from a compact source, like a planet or a star: see \cite{Calderon18} and references therein for a review of different possibilities.  In this paper we consider a formation channel where a close encounter between a red giant and a more compact star leads the compact star to accrete a tenuous envelope.

In this article we develop a framework that allows dynamical simulations of nuclear star cluster formation to be post-processed to take into account the effects of stellar collisions and close encounters on the stellar populations stars present at the Galactic Centre.  In order to be computationally tractable, such simulations employ superparticles that represent many stars: in our case $200\,\msun$ of stars per superparticle.  We attach a stellar population to each superparticle and follow the effects both of stellar evolution and stellar collisions on the population.  We obtain the local stellar number density and velocity dispersion from the dynamical simulation and use it to calculate the collision rate as a function of time separately for each superparticle.  The effects of the collisions on the pair of stars that collide are derived from a combination of physical reasoning and parameterised fits to hydrodynamical simulations of stellar interactions.

The remainder of this article is organised as follows. Sections~\ref{sect:nbody} and~\ref{sect:coagulator} describe the contents of our models: $N$-body simulations of the formation of the NSC and the treatment of stellar collisions and close encounters.  Section~\ref{sect:effectsOfEncounters} describes the effects of encounters on the stellar populations in the NSC, and Section~\ref{sect:fuzzballs} the behaviour of compact stars that have accreted a tenuous envelope.  Following a discussion in Section~\ref{sect:discussion} we conclude in Section~\ref{sect:conclusions}.

\section{$N$-body simulations}
\label{sect:nbody}
The $N$-body run used in this paper is described in \cite{Antonini12}, \cite{Mastrobuono14}, \cite{PMB14} and \cite{Tsatsi17}.  In these works we modelled the formation of a Milky Way-like NSC, through the inspiral and merger of twelve dense and massive stellar clusters, similar to globular clusters (GCs), in the centre of a nuclear bulge of mass $10^8\,\msun$. 
The Galactic centre initially hosts a supermassive black hole (SMBH) of $4\times10^6\,\msun$, similar to Sgr A$^*$ \citep{GH98, EI05, GI09, BGS16, GP17}. 
All clusters are described by a tidally truncated \cite{King66} model with total mass of $1.1\times10^6\,\msun$, core radius $r_c=0.5$\,pc and concentration parameter $W_0=5.8$. 
The adopted simulation corresponds to `Simulation 1' in \cite{Tsatsi17}.  While the nuclear bulge is modelled using $N_b=227523$ equal mass ($400\,\msun$) $N$-body particles, every GC is modelled using $N_{GC}=5715$ superparticles of $200\,\msun$. We assume that the star clusters have already migrated to the inner Galaxy and that each cluster is initially located $20$\,pc from the central SMBH. All GCs are on randomly inclined circular orbits.

Our GCs decay consecutively at regular intervals of time ($0.85$\,Gyr), over $\sim10.2$\,Gyr. The system is then left to evolve without any additional GC infall until it reaches an age of $12.4$\,Gyr. At the end of the simulation, we obtain a central NSC that kinematically (including its rotation and velocity dispersion) and morphologically (e.g. mass, shape and density profile) resembles the one observed at the centre of the Milky Way. 

\subsection{Scaling}
\label{sect:scaling}
To obtain a more accurate match between the final mass and central density of the simulated and observed NSCs, we scaled our system using the observed values of the half-mass radius and total mass for the Galactic nucleus \citep{Schoedel14a,Schoedel14b}.

We defined the scaling factor for the distances as the ratio between the observed projected half-light radius ($R_{\rm h,obs}=4.2$\,pc) and the simulated projected half-mass radius $R_{\rm h,sim}$. Following \citet{Schoedel14a} we use the projected radial co-ordinate
\begin{equation}
    R = \sqrt{x^2 + (y/q)^2}
\end{equation}
which accounts for the oblateness $q$.  The observed NSC has $q=0.64$ whilst our model has $q=0.72$ \citep{Tsatsi17}. We obtain
\begin{equation}
    l_{\rm scale}=R_{\rm h,obs}/R_{\rm h,sim} = 0.448
\end{equation}
The mass scale is the ratio between the observed NSC mass, $M_{\rm obs}=2.5\times10^7\,M_\odot$, and the simulated NSC mass $M_{\rm sim}$:
\begin{equation}
    m_{\rm scale}=M_{\rm obs}/M_{\rm sim} = 1.82
\end{equation}

Finally, the velocities have been scaled so as to keep the virial radius $Q$ of the simulated NSC constant after the other two scalings have been applied.  Since
\begin{equation}
    Q = \frac{2T}{V} = {\sum_im_iv_i^2} / {\sum_{ij}\frac{m_im_j}{\left|\mathbf{r}_i-\mathbf{r}_j\right|}}
\end{equation}
where $m_i$ and $v_i$ are the superparticle masses and velocities, and $\mathbf{r}_i$ their positions, we obtain the velocity scaling
\begin{equation}
    v_{\rm scale} = \sqrt{\frac{m_{\rm scale}}{l_{\rm scale}}} = 2.02.
\end{equation}

The scaled number density is then defined as
\begin{equation}
    n'=n/l_{\rm scale}^3,
\end{equation}
and the spatial and projected mass densities are
\begin{equation}
    \rho'=\rho\times m_{\rm scale}/l_{\rm scale}^3
\end{equation}
and
\begin{equation}
    \Sigma'=\Sigma\times m_{\rm scale}/l_{\rm scale}^2.
\end{equation}

\begin{figure}
    \begin{center}
        \includegraphics[width=\columnwidth]{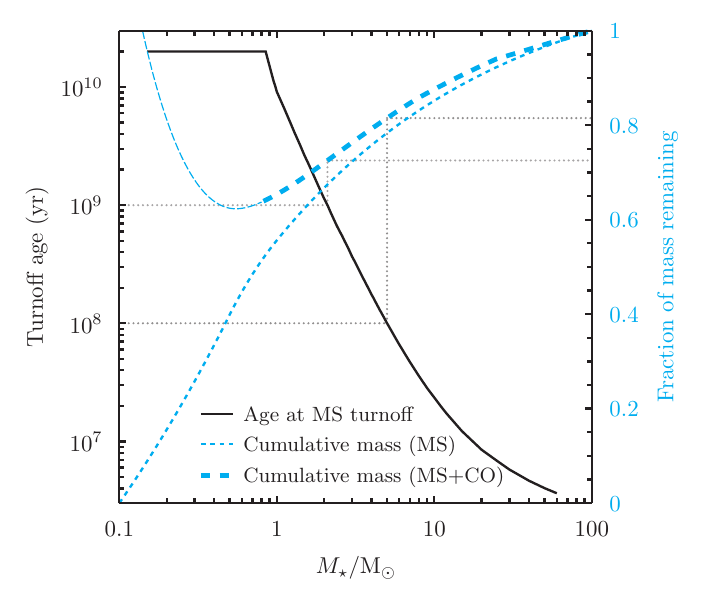}
        \caption{Turnoff age (left-hand axis, black solid line) and fraction of population mass remaining (right-hand axis, blue lines) as functions of main-sequence turnoff mass $M_{\star}$.  The short-dashed line only includes main-sequence stars; the long-dashed line includes the mass of compact remnents.  The thin segment of the long-dashed line is unphysical owing to our assumption of a constant white dwarf mass.  The grey dotted lines show how to interpret the graph to calculate the fraction of stellar mass remaining at $10^8$ and $10^9$ years.}
        \label{fig:massEvolution}
    \end{center}
\end{figure}

It is also necessary to consider how to treat the evolution of the particle masses.  In the $N$-body code, the superparticles have a constant mass, but when we include the effects of stellar collisions and stellar evolution, the mass changes as the stellar population evolves.  Essentially it is necessary to choose a time to set the $N$-body superparticle masses and the  mass in the code where we evolve the stellar population equal.  Figure~\ref{fig:massEvolution} shows the range of options available.  The black line, plotted against the left-hand ordinate axis, shows the turnoff age in our models as a function of turnoff mass, whereas the blue curves (plotted against the right-hand ordinate axis) show the stellar mass at that age as a fraction of the initial population mass.  Grey dotted lines show that the fractions of mass remaining at $10^8$ and $10^9$ years are 0.815 and 0.725 respectively.  We choose to set the masses equal at $10^8\,{\rm yr}$, since although that time precedes most of the dynamical evolution in the $N$-body code, a large fraction of the mass loss has already happened.  Hence we define a second mass scale as the ratio between the initial stellar mass in each superparticle and the simulated mass, $m_{\rm scale,0}=2.23$.

%We note that the virial ratio of the system is $Q=1.47$.

%\begin{figure*}
%    \centering
%    \includegraphics[width=0.45\textwidth]{3D_density}~\includegraphics[width=0.45\textwidth]{Surface_density}
%    \caption{Spatial (left panel) and projected (right panel) density profile of the rescaled NSC.}
%    \label{fig:density}
%\end{figure*}

\subsection{Dynamical interactions}
In order to make a rough estimate of the degree of dynamical interaction between stars we introduce the {\it interaction parameter} for $N$-body superparticle $i$, $\Gamma_{\odot,i}$, defined as

\begin{equation}
\Gamma_{\odot,i} = \int4\sqrt{\pi}\sigma_i(t) n_i(t) R_{\odot}^2\left(1+\frac{2GM_{\odot}}{2\sigma_i(t)^2R_{\odot}}\right)\ud t
\end{equation}
where $n_i(t)$ is the number density of stars in the particle $i$ and $\sigma_i(t)$ is the velocity dispersion local to the particle.  The integral is carried out over all times from the introduction of the particle to the simulation until the end of the simulations at 12.4\,Gyr.  $\Gamma_\odot$ can be understood as the collision probability for an individual star within the $N$-body particle, if it had at all times the mass and radius of the Sun.  Whilst the actual collision probability for any given star must be calculated by a full integral over the time-varying stellar population, as described below, this gives a simple way of assessing how much interaction the stars that make up a given $N$-body particle are likely to experience.

\begin{figure}
    \begin{center}
        \includegraphics[width=\columnwidth]{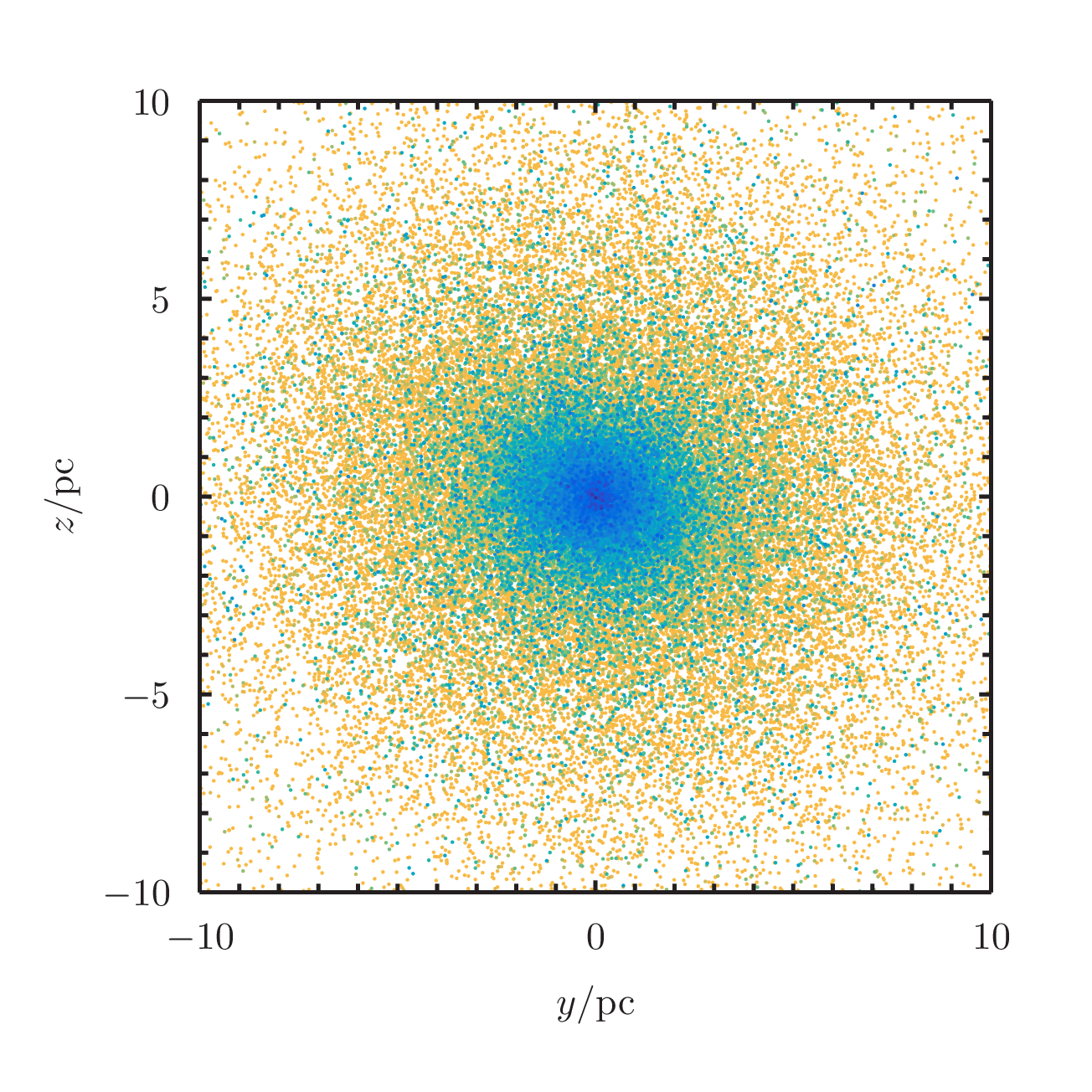}\\
        \includegraphics[width=\columnwidth]{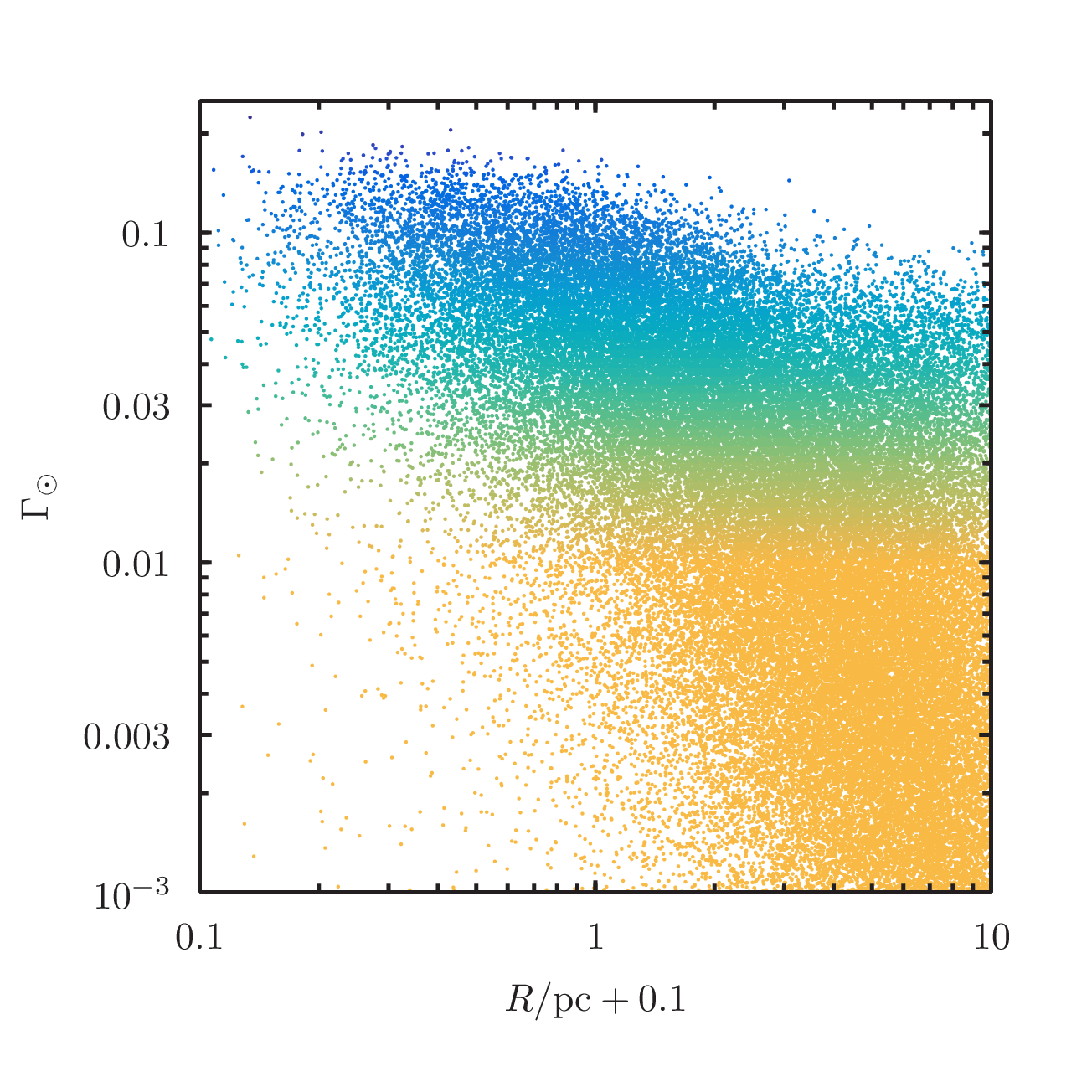}
    \end{center}
    \caption{{\it Top panel:} Cluster snapshot, projected along the line of sight, in pc.  Particles are coloured according to $\Gamma_\odot$, plotted so that the most interactive particles are on top.  {\it Bottom panel:} the interaction parameter $\Gamma_\odot$ is plotted as a function of projected galactrocentric radius $R$.  An offset of 0.1\,pc is added to make the plot more compact.  The colour scale is the same as in the top panel.}
    \label{fig:snapshot}
\end{figure}

The final positions of the $N$-body superparticles, projected onto the $y-z$ plane, are shown in Figure~\ref{fig:snapshot}.  The colour scheme tracks the value of $\Gamma_\odot$, as can  be seen in the bottom panel.  Whilst the particles are plotted in an order that systematically highlights the most interactive particles, it can be seen that these all lie towards the centre of the cluster, with the outer parts comprising entirely non-interactive particles.  Hence we expect that the effects of close encounters and collisions will be most pronounced at the centre of the cluster.

\section{Treatment of stellar collisions and close encounters}
\label{sect:coagulator}

In order to make a quantitative calculation of the effects of stellar evolution, collisions and close encounters on the stellar population in the Milky Way NSC, we attach a population of stars to each superparticle, calculate the rates of collisions between the various stellar species, and model the effects of the collisions that occur.  The populations are coeval and each have initial mass 
\begin{equation}
    M_{\rm tot,0} = f_{\rm ow}M_{\rm sp}m_{\rm scale,0}
\end{equation}
where the superparticle mass in the $N$-body simulations, $M_{\rm sp}=200\,\msun$ (see Section~\ref{sect:nbody}) and $m_{\rm scale,0}$ is described in Section~\ref{sect:scaling}.  The additional scale $f_{\rm ow}$ allows us to over-sample the IMF in order to avoid introducing problems from stochastic sampling.  Empirically we have found that $f_{\rm ow}=10$ is sufficient, which corresponds to one or two black holes per superparticle.  The stars are all taken to be single, and their birth masses of the stars are distributed according to the IMF of \citet{KTG93}.

Having built a stellar population, we evolve it using the dynamical history of that superparticle from the $N$-body simulations.  We take the number density and velocity dispersion of the 100 superparticles closest to the particle being modelled and use that to compute the stellar interactions using the stellar population attached to the given superparticle.  This effectively requires us to assume that the stellar population is only weakly affected by stellar encounters, an assumption that turns out to be quite good.

The number density $n_i$ of species $i$ is then given by
\begin{equation}
    n_i = n_{\rm sp}N_i/f_{\rm ow}
\end{equation}
where $n_{\rm sp}$ is the number density of superparticles local to the superparticle being modelled and $N_i$ is the number of stars of species $i$ in the simulation.  \\
The velocity dispersions are taken to be Maxwellian, and the velocity dispersion $\sigma$ is taken directly from an average over the 100 nearest superparticles in the $N$-body simulations.

The instantaneous collision rate of an individual star $i$ with mass $M_i$ and radius $R_i$ with a star of type $j$ at velocity $v_\infty$ is then given by \citep[][Eqn.~8-116]{BinneyTremaine87}

\begin{equation}
\mathcal{R}_{ij} = \int \mathcal{R}^v_{ij}(v_\infty)f(v_\infty)\ud v_\infty,
\end{equation}
where
\begin{equation}
\mathcal{R}^v_{ij}(v_\infty) = n_j \uppi R_{ij}^2 v_\infty \left[1+\frac{2G(M_{i}+M_j)}{v_\infty^2 R_{ij}}\right]
\label{eqn:gammarv}
\end{equation}
and $R_{ij} = R_i + R_j$.  For cases where the outcome is largely insensitive to $v_\infty$, such as the collision of two main-sequence stars, we normalise over the Maxwellian velocity distribution to obtain
\begin{equation}
\mathcal{R}_{ij} = 4\sqrt{\pi}\sigma n_j R_{ij}^2\left[1+\frac{G(M_i+M_j)}{2\sigma^2R_{ij}}\right].
\end{equation}
For collisions involving red giants, where in general both the velocity and pericentre separation of the collision make a large difference to the outcome we first sample $v_\infty$ from the velocity distribution.  There is a physical collision in timestep $\Delta t$ when $X<\mathcal{R}^v_{ij}\Delta t$, where $X$ is uniformly chosen between 0 and 1, and $\mathcal{R}^v_{ij}$ comes from Eqn.~\ref{eqn:gammarv}. We obtain the collision radius as
\begin{equation}
R_{\rm col} = \frac{R_{\rm esc}}{2}\left[\sqrt{1+4\left(\frac{R_{\rm rand}}{R_{\rm esc}}\right)^2}-1\right]
\end{equation}
with $R_{\rm esc}=2G(M_i+M_j)/v_\infty^2$ and $R_{\rm rand}^2=X/(\uppi n_j v\Delta t)$. 

In order to accelerate the calculation we bin the stars by mass, and treat all the main-sequence stars in a given bin as being equivalent.  This is possible since the stellar radius and mass do not change very much over the main sequence for the low-mass stars that are relevant for our analysis.  The main-sequence lifetime of each star is calculated individually, however, depending on its precise mass, and we treat giants individually since their masses and radii change significantly.  Once the stars have evolved into compact remnants we again bin them, and have only a single compact object species of each type.  For convenience the bin masses are the same as the masses for which stellar evolution tracks are calculated, as described below.

\subsection{Stellar evolution}

We follow the evolution of single stars using the stellar evolution code {\sc stars}, originally written by \citet{Eggleton71}, using the physical prescriptions described by \citet{Pols+95}.  Details of modifications to the code to make it run autonomously can be found in \citet{Church06}, but are not particularly significant for this work.  Our evolution tracks are computed at solar metallicity, with wind mass loss on the first giant branch and during core helium burning given by the formula of \citet{Kudritzki+Reimers78}, with $\eta=0.4$.  On the asymptotic giant branch we follow the formula of \citet{VassiliadisWood93}.  Our mass-loss law is not particularly suitable for massive stars, but our $N$-body tracks start at a time of $\approx 400\,{\rm Myr}$, by which time all of the massive stars have already evolved into compact objects.

Instead of calculating the evolution of each star individually, we make models of single stars at 76 different masses, chosen carefully to resolve the physics relevant to the questions that we are trying to answer.  For each star of initial mass $m_{i,0}$ we define a set of times $t_{j}(m_{i,0})$ which delimit key phases in the evolution of the star (end of core H burning, base of giant branch, etc.).  The list of key times and their definitions is given in Table~\ref{tab:keyPoints}

\begin{table}
    \begin{tabular}{lll}
        \hline
        $j$     & Point     & Definition    \\
        \hline
        1   & End of MS     & Core H mass fraction reaches zero \\
        2   & Base of RGB   & Minimum in surface luminosity \\
        3   & Start of HB   & Surface luminosity peak at RGB tip\\
        4   & End of HB     & Core He mass fraction reaches zero \\
        5a  &               & Exhaustion of H and He burning \\
        \ b & CO formation  & Code non-convergence during C burning\\
        \ c &               & Nuclear reactions beyond C burning \\
        \hline
    \end{tabular}
    \caption{Key points in the evolution of the stars and how they are defined}
    \label{tab:keyPoints}
\end{table}

To compute the state of a star of initial mass $m_0$ at time $t$ we then use the following procedure:
\begin{enumerate}
    \item Identify the two tracks whose initial masses $m_{i,0}$ and $m_{i+1,0}$ bracket $m_0$.
    \item Interpolate the log of the key times, $\log t_{j}$, linearly\footnote{It is necessary for the interpolations to be linear since the gradients of e.g. $t_{j}$ with mass do not vary smoothly around discontinuities in the stellar physical properties, for example the appearance of the core convection zone.} in mass between the two tracks to obtain a set of key times for the star in question, $t_{j}(m_0)$.
    \item Identify the pair of key times $t_j(m_0)$ and $t_{j+1}(m_0)$ that bracket the current time $t$ and calculate the fractional time 
        \begin{equation}
            \tau = \frac{t-t_j(m_0)}{t_{j+1}(m_0)-t_{j}(m_0)}.
        \end{equation}
    \item Identify the corresponding times in the two bracketing tracks; i.e. 
        \begin{equation}
            t_{{\rm interp},i} = t_j(m_{i,0}) + \tau \times \left[t_{j+1}(m_{i,0}) - t_{j}(m_{i,0})\right]
        \end{equation}
        and similarly for $m_{i+1,0}$, and interpolate each stellar property (mass, core mass, $\log T_{\rm eff}$, etc.) to that time.
    \item Finally interpolate each stellar property linearly in mass between the two bracketing tracks at the corresponding values of $t_{\rm interp}$.
\end{enumerate}

This complicated procedure allows us to accurately reproduce stellar behaviour over the full range of masses using 76 tracks; a spot check of interpolated tracks against generated ones shows essentially identical HR diagrams with typical time-scale errors of $\approx 0.02$ per cent.  We choose this process in preference to using the fitting formulae of \citet{Hurley+00} because we wish to use additional derived quantities from the full models in the case of stellar mergers.  To obtain $K$-band magnitudes we use the bolometric corrections and empirically calibrated $T_{\rm eff}$-colour relations of \citet{Lejeune+98}.

At the end of its nuclear burning lifetime each remaining star is converted into a compact remnant.    For performance reasons, we consider only four species of compact remnant; white dwarfs produced by normal single-star evolution (WDs), white dwarfs that have been produced by stellar collisions stripping the envelope of giant stars (cWDs), neutron stars (NSs), and black holes (BHs).  Initial stellar mass ranges and compact object masses for these species are listed in Table~\ref{tab:COs}.

\begin{table}
    \begin{tabular}{lrr}
        \hline
        CO species &   Initial mass range  & CO mass \\
        \hline
        WD          & $<8\,\msun$           & $0.55\,\msun$ \\
        cWD         &                       & $0.5\,\msun$  \\
        NS          & $8$ -- $25\,\msun$    & $1.4\,\msun$  \\
        BH          & $>25\,\msun$          & $10\,\msun$   \\
        \hline
    \end{tabular}
    \caption{Initial mass ranges and compact object masses for each compact object (CO) species in our stellar evolution model.}
    \label{tab:COs}
\end{table}

\subsection{Treatment of collisions and close encounters}
\label{sect:collisionTreatment}

We model collisions of both types of evolving star (MS and RG) with all types of compact object (cWD, WD, BH and NS), as well as collisions of MS stars with other MS stars and RGs.  A summary can be found in Table~\ref{table:colmatrix}.  In the following subsections we describe and justify the collision prescriptions that we implement.

\begin{table*}
    \begin{tabular}{lp{10em}p{10em}p{10em}p{10em}}
      & \multicolumn{4}{c}{Impactor species} \\
      & \multicolumn{4}{c}{$\overbrace{\hspace{45em}}$} \\
    Impacted & Main sequence & Red giant & Light    compact object   & Black hole \\
    species  & MS            & RG        & cWD / WD / NS             & BH        \\
\hline
    MS      & Merger        & Fuzzball formation    & MS destroyed & MS destroyed \\
    RG      &  Mass loss    & Ignored               & Mass loss from giant & Mass loss from giant \\
    cWD/WD/NS & No effect   & Fuzzball formation    & Ignored       & Ignored \\
    BH      &  No effect    & Fuzzball formation    & Ignored       & Ignored \\
\hline 
    \end{tabular}
    \caption{Effect of impactor (columns) on species collided with (rows).  The table is not symmetric since a collision between two stars of different types will have different effects on each.  See Section~\ref{sect:collisionTreatment} for details}
    \label{table:colmatrix}
\end{table*}

\subsubsection{Main-sequence -- main-sequence collisions}
\label{sect:MSMS}

We assume that any collision between two main-sequence stars leads to merger with no loss of mass.   Velocity dispersions in our model of the Galactic Centre are typically between 100 and 300$\,\kms$, which is comfortably less than the surface escape velocities of main-sequence stars ($\approx600\,\kms$ for the Sun).  This treatment ignores the small fraction of mass that is expected to be lost in such encounters, and the high-velocity tail of potentially destructive encounters; neither are expected to be significant for our results.

% {\ri XXX Add ref to Frietag et al. collision simulations}

\subsubsection{Main-sequence -- red giant collisions}
\label{sect:MSRG}

For red giants, the analysis of the previous section is reversed.  For a $100\,\Ro$, $1\,\msun$ giant, the surface escape velocity $v_{\rm esc,RG}\approx60\,\kms$, less than the typical encounter velocity at the Galactic centre.  The dwarf is also very much more dense than the outer layers of the giant.  As a result, the most likely outcome is for the dwarf to pass unscathed through the outer layers of the giant's atmosphere.  Some of the giant's envelope is removed by the bow shock driven by the impacting main-sequence star, and some of that lost mass is accreted by the impactor.

\begin{figure*}
\includegraphics[width=\textwidth]{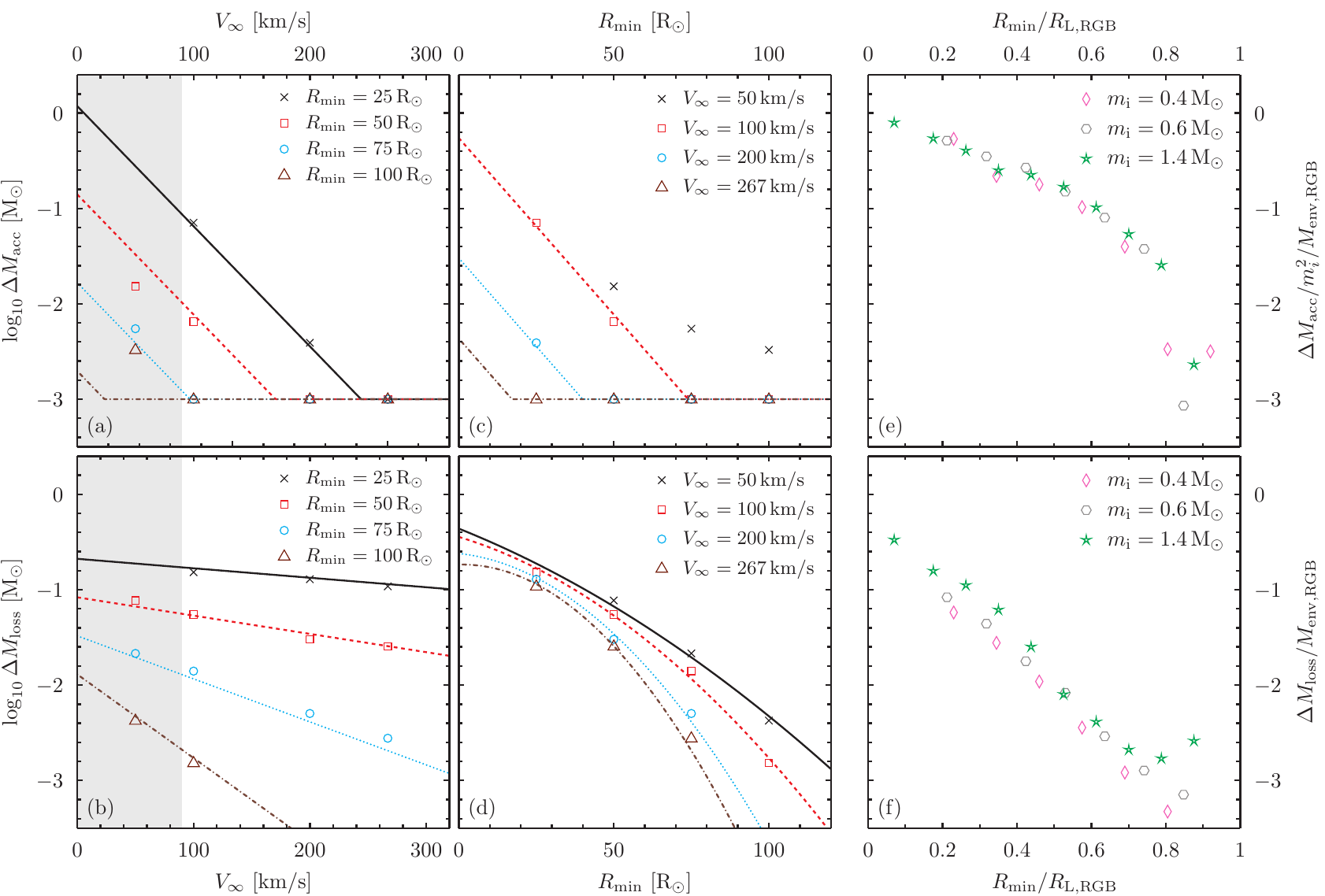}
\caption{{\it Panels (a) to (d):} Empirical fits for the mass lost to infinity (bottom panels) and accreted onto the impactor (top panels) for a collision of a $2\,\msun$ giant with a $1\,\msun$ point-mass impactor.  Data are taken from figure 6 of BD99.  The left panels show the variation of $\Delta M$ with relative velocity at infinity $V_\infty$, plotted for four different values of the pericentre separation $R_{\rm min}$: $25\,\Ro$ (black crosses, solid black fit line), $50\,\Ro$ (red squares, red dashed fit line); $75\,\Ro$ (blue circles, blue dotted fit line), and $100\,\Ro$ (brown triangles, brown dot-dashed fit line).  The right panels show the variation of $\Delta M$ with $R_{\rm min}$, plotted for four different values of $V_\infty$: $50\,\kms$ (black crosses, solid black fit line),  $100\,\kms$ (red squares, red dashed fit line); $200\,\kms$ (blue circles, blue dotted fit line), and $267\,\kms$ (brown triangles, brown dot-dashed fit line).  The grey shaded area shows the region where $V_\infty < V_{\rm esc,RG}$. {\it Panels (e) and (f):} Investigation of the scaling of the mass lost to infinity (f) and accreted by the point mass (e) in encounters between a $0.8\,\msun$ giant and three different point-mass impactors.  The impactors had masses of $0.4\,\msun$ (pink diamonds), $0.6\,\msun$ (grey hexagons) and $1.4\,\msun$ (green stars).  Results are taken from \citet{DaviesBenzHills91} and \citet{DaviesBenzHills92}.  In panel (e) we plot the  mass accreted by the impactor, $\Delta M_{\rm acc}$, divided by $M_{\rm env,RGB}m_{\rm i}^2$, where $M_{\rm env,RGB}$ is the mass of the giant envelope and $m_{\rm i}$ is the mass of the point-mass impactor.  The three sets of points lie on a single trend with $_{\rm min}/R_{\rm L,RGB}$, which suggests that this is the correct scaling of the accreted mass with mass of the giant and impactor, and closest approach distance $R_{\rm min}$.  Similarly, in panel $(f)$ the mass lost by the giant, $\Delta M_{\rm loss}$ is divided by $M_{\rm env,RGB}$ and again shows a single trend.}
\label{fig:BD99fits}
\end{figure*}

We parameterise these effects following the simulations presented by \cite{BaileyDavies99} (BD99 hereafter).  They studied collisions between a $2\,\Mo$ giant and a $1\,\Mo$ main-sequence star.\footnote{Whilst BD99 in fact studied a much larger range of impactor masses and two different target giants, they only presented detailed results for this one case, and the original simulation outputs are no longer available.}  They present measurements of the mass lost from the system and the mass accreted by the impactor as a function of the encounter pericentre $R_{\rm min}$ and velocity at infinity $v_\infty$ (figure 6 of BD99).  We find that, when $v_\infty > v_{\rm esc,RG}$, their results are well fit by the empirical formulae

\begin{equation}
% \log_{10}\frac{2\Delta M_{\rm lost}} {M_{\rm RG}}=a_R + b_R \r + (a_{RV} + b_{RV} \r + c_{RV} \r^2)\v
\Delta M_{\rm lost,BD99} = \frac{M_{\rm RG}}{2\,\Mo}10^{f_{\rm loss}(\r,\vt)}
\end{equation}
where
\begin{equation}
f_{\rm loss}(\r,\vt)=a_R^{\rm loss} + b_R^{\rm loss} \r + (a_{RV}^{\rm loss} + b_{RV}^{\rm loss} \r + c_{RV}^{\rm loss} \r^2)\vt,
\label{eqn:MSRGB-Mloss}
\end{equation}
and
\begin{equation}
%\log_{10}\frac{2\Delta M_{\rm acc}} {M_{\rm RG}}= \max(-3, a_R + b_R \r + b_{V} \v)
\Delta M_{\rm acc,BD99} = \frac{M_{\rm RG}}{2\,\Mo}10^{f_{\rm acc}(\r,\vt)},
\end{equation}
where
\begin{equation}
f_{\rm acc}(\r,\vt) = \max(-3, a_R^{\rm acc} + b_R^{\rm acc} \r + b_{V}^{\rm acc} \vt),
\label{eqn:MSRGB-Macc}
\end{equation}
$\r=R_{\rm min}/R_{\rm RG}$, and $\vt=V_\infty/V_{\rm esc,RG}$. The coefficients are given in Table~\ref{table:MSRGcoeffs} and the fits are plotted graphically in Figure~\ref{fig:BD99fits}. 

\begin{table}
\begin{tabular}{llllll}
\hline
\multicolumn{6}{l}{Mass loss (RGB -- MS/WD/NS)}\\
Coefficient     &$a_R^{\rm loss}$  &$b_R^{\rm loss}$  &$a_{RV}^{\rm loss}$   &$b_{RV}^{\rm loss}$   &$c_{RV}^{\rm loss}$   \\
Value           & -0.27 & -1.52 & -0.16     & 0.54      & -1.06     \\
\hline
\multicolumn{6}{l}{Accreted mass (RGB -- MS/WD/NS)}\\
Coefficient     &$a_R^{\rm acc}$  &$b_R^{\rm acc}$  &$b_V^{\rm acc}$      \\
Value           & 0.998 & -3.48 & -1.13     \\
\hline
\multicolumn{6}{l}{Mass Loss (RGB -- BH)}\\
Coefficient     & $a_v^\bullet$ & $b_v^\bullet$ & $\r_{\rm crit}$ \\
Value           & 0.82          & 0.20          & 0.32 \\
\hline
\end{tabular}
\caption{Values of the coefficients in Equations~\ref{eqn:MSRGB-Mloss}, \ref{eqn:MSRGB-Macc} and \ref{eqn:MlossD09}}
\label{table:MSRGcoeffs}
\end{table}

As we only have data for one giant mass, it is necessary to scale the results to other stellar masses.  Physical reasoning suggests that, for homologous giant structures, the total mass and surface escape velocity are reasonable quantities to scale the mass loss with. 
It is also necessary to generalise the results from BD99 to different impactor masses.  A more massive impactor will affect the atmosphere of a giant star from further away, and will both eject and accrete more mass as it passes through.  In the absence of a comprehensive set of simulations  we make two assumptions.  Firstly, since the mass loss from a giant in an encounter is somewhat analogous to tidal stripping, instead of scaling $R_{\rm min}$ with the giant radius we scale with the giant's instantaneous Roche lobe radius at pericentre; i.e. with $R_{\rm RG}r_L(q)$ where the mass ratio $q=M_{\rm RG}/m_{\rm i}$ for impactor mass $m_{\rm i}$ and
\begin{equation}
r_{\rm L}(q) = \frac{0.49q^\frac{2}{3}}{0.6q^{\frac{2}{3}} + \log(1+q^\frac{1}{3})}
\end{equation}
following \citet{Eggleton83}.  Secondly, we assume that the accretion of mass by the impactor as it passes through the giant's envelope largely follows the mechanism of \citet{BondiHoyle44} and hence, following their equation 1, should be scaled with $m_{\rm i}^2$.

To test these assumptions we utilise the results of \citet{DaviesBenzHills91} and \citet{DaviesBenzHills92}.  They made SPH models of interactions of a $0.8\,\msun$ giant with $0.4$, $0.6$ and $1.4\,\msun$ point masses.  Whilst they took $V_\infty=10\,\kms$, appropriate to globular clusters but not to the majority of encounters at the Galactic Centre, the impactor velocities inside the giant envelope are still well in excess of the local sound speed, and so the relevant physics should not be very different.  The right-most panels of Figure~\ref{fig:BD99fits} show their results scaled as described in the paragraph above; the curves for the three different impactor masses show rough agreement other than at very low masses where the simulations are underesolved.  We consider that this validates our assumptions in the absence of further hydrodynamic simulations.  Our final formulae for the mass lost and accreted by the impactor in giant--dwarf encounters are therefore

\begin{equation}
\Delta M_{\rm lost} = \frac{M_{\rm RG}}{2\,\Mo}10^{f_{\rm loss}(\r,\vt)}
\end{equation}
and
\begin{equation}
\Delta M_{\rm acc} = \left(\frac{m_i}{\msun}\right)^2\frac{M_{\rm RG}}{2\,\Mo}10^{f_{\rm acc}(\r,\vt)}
\end{equation}
with $f_{\rm loss}$ and $f_{\rm acc}$ given by Equations~\ref{eqn:MSRGB-Mloss}~and~\ref{eqn:MSRGB-Macc}, and
\begin{equation}
\r=R_{\rm min}/R_{\rm RG}\frac{r_{\rm L}(q)}{r_{\rm L}(2)}.
\label{eqn:rtwiddle}
\end{equation}

For each giant in our simulations, we accumulate the mass lost in successive collisions with main-sequence stars and compact objects.  At the point when the total mass lost in collisions is greater than the remaining envelope mass the giant is converted into a compact object; i.e. even weak collisions that do not remove the entire envelope can shorten the giant's lifetime.  The second-order effect of the change in stellar evolution induced by the lower envelope mass is ignored.

For the main-sequence star, we assume that the (relatively compact) main-sequence star acquires a hot, low-density halo of accreted material, as seen in hydrodynamic simulations of the encounters of compact stars with giants.  We refer to such objects as {\it fuzzballs}.  The amounts of mass accreted are typically rather small and we do not expect them to significantly affect the long-term evolution of the main-sequence stars, but we record the formation of these objects and consider their formation rates and expected properties further in Section~\ref{sect:fuzzballs}.

\subsubsection{Compact object -- red giant collisions}

For low-mass compact objects (WDs, cWDs and NSs) colliding with giants, our treatment follows that for MS--RG collisions in Section~\ref{sect:MSRG} above.  The density contrast between a giant envelope and a main-sequence star is so large that the impactor can be treated as a point mass (and indeed is in the SPH simulations that we base our approach on).  The same treatment then obviously applies for compact objects.  Such collisions will also produce fuzzballs by accretion from the giant envelope.

\begin{figure}
    \begin{center}
        \includegraphics[width=.8\columnwidth]{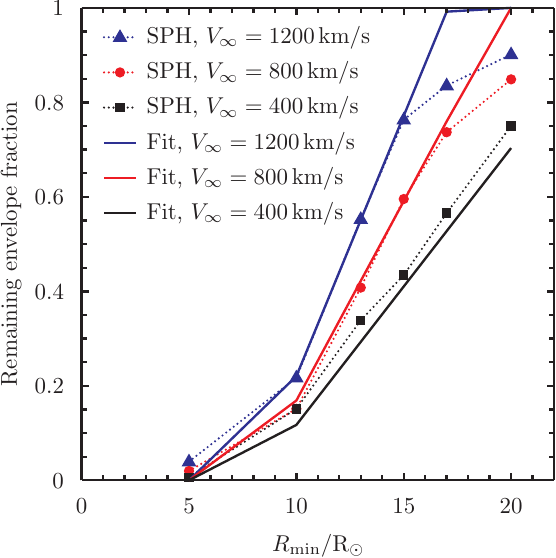}
    \end{center}
    \caption{Fraction of the stellar envelope remaining after collisions between a $1\,\msun$ giant and a $10\,\msun$ black hole.  The giant has radius $25\,\Ro$.  Points with dotted lines show results from the SPH simulations of \citet{Dale+09}: black squares are for $V_\infty=400\,\kms$, red circles for $V_\infty=800\,\kms$ and blue triangles $V_\infty=1200\,\kms$.  Solid lines with matching colours show our fits for the corresponding velocity values.}
    \label{fig:fitDale}
\end{figure}

For collisions of black holes with giants, the mass ratio, and hence behaviour, is typically rather different.  We base our treatment on \citet{Dale+09} and in particular their figure~11, where they present the envelope fraction remaining after collisions of a $1\,\msun$ red giant with a $10\,\msun$ black hole.  We fit their results as
\begin{equation}
\Delta M_{\rm lost} = (M-M_{\rm core})\max\left(0,\left[1-f_{\rm ret}(\r,\vt)\right]\right),
\label{eqn:MlossD09}
\end{equation}
where
\begin{equation}
f_{\rm ret}(\r,\vt)  = \left\{
\begin{array}{ll}
0   & \r<\r_{\rm crit}\\
\left(a_v^\bullet + b_v^\bullet\vt\right)(\r-\r_{\rm crit}) & \r > \r_{\rm crit}\\
\end{array}
\right.
\end{equation}
and, as before, $\vt=V_\infty/V_{\rm esc, RG}$, $\r=R_{\rm min}/R_{\rm RG}\frac{r_{\rm L}(q)}{r_{\rm L}(10)}$ analogously with Equation~\ref{eqn:rtwiddle}, and the fit coefficients are given in Table~\ref{table:MSRGcoeffs}.  A comparison of the SPH results and fit is plotted in Fig.~\ref{fig:fitDale}: whilst the fit is not perfect it provides a conservative estimate of the effect of such collisions on the population of red giants.

Most collisions that red giants suffer with other stars remove only part of their envelope.  \citet{Dale+09} investigated the post-collision evolution of giants and found that it was necessary to remove almost all of the envelope to significantly affect the visibility of the giants.  As with MS--RG collisions, for encounters that do not remove all of the envelope, we keep a running sum of the mass removed, which we add to the mass loss through stellar winds that the giant would have experienced in any case.  When this total exceeds the mass of the envelope we take the giant to have evolved into a compact object.  In this manner we are able to calculate a history for each giant without resorting to carrying out prohibitively expensive hydrodynamic or stellar evolution calculations on a star-by-star basis.

\subsubsection{Main sequence -- compact object collisions}

We expect collisions between MS stars and compact objects to usually result in capture of the compact object by the star, following the same reasoning as in Section~\ref{sect:MSMS}.  Following capture, the accreted compact object will be more dense than the rest of the star, and hence will sink into the star's core.  This will lead to the formation of a bright giant.  If the compact object is a white dwarf the giant will be very similar to an ordinary star at the tip of the AGB, if it is a neutron star a Thorne-\.Zytkow object \citep{Thorne75,Thorne77} will form, and if the compact object is a black hole then the result will be a quasistar \citep{Begelman06, Begelman08,Volonteri10}.  The details of the evolution of these objects are complicated and uncertain, but for us the most significant property is that they are expected to be short-lived.  The MS star is thus effectively destroyed, and the compact object recovered as the final product of the evolution of the bright giant.  It is possible that e.g. a neutron star could be converted into a black hole by this process, but it is rare enough that we do not need to take this pathway into account.

\subsubsection{Collisions between pairs of giants}

Collisions between pairs of giants could be as destructive as collisions between giants and main-sequence stars of similar mass, or perhaps even more so given the lower density contrast.  However they are rare.  Making a simple comparison between the rate of collision of two identical giants of mass $M$ and radius $R_{\rm g}$, and the rate of collision of a giant with a main-sequence star of equal mass $M$, we find that
\begin{equation}
\frac{\mathcal{R}_{\rm giant}}{\mathcal{R}_{\rm MS}} = 4\frac{n_{\rm giant}}{n_{\rm MS}} \left[1+\frac{GM}{2\sigma^2R_{\rm g}}\right]\left[1+\frac{GM}{\sigma^2R_{\rm g}}\right]^{-1}
\end{equation}
an enhancement of a factor of between two and four, depending on the velocity dispersion, from the ratio of number densities.  But the giant phase is short, so the number density ratio is always small, particularly when lower-mass main-sequence stars are taken into account.  Therefore we choose to ignore giant-giant collisions.

\subsubsection{Collisions between pairs of compact objects}

Collisions between pairs of compact objects are very rare owing to the small radii of the objects.  Their only effect on our work would be if they removed a significant fraction of the compact objects, but this is not expected.  We ignore them entirely \citep[see e.g. ][for an analysis of collisions between compact objects in galactic nuclei]{Quinlan90, Benz89}.
%#########################################################################################

\section{Effects of encounters on stellar populations}
\label{sect:effectsOfEncounters}
\begin{figure}
    \includegraphics[width=\columnwidth]{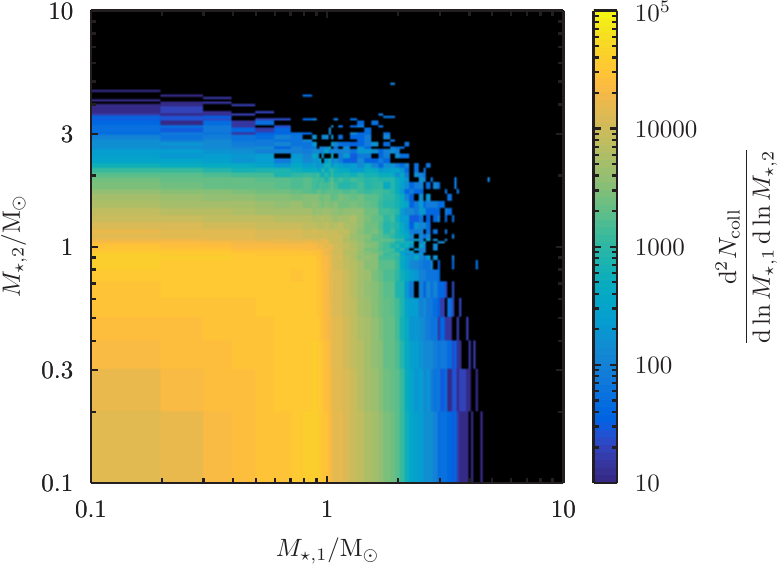}
    \caption{Relative magnitude of the effect of collisions on stars of different masses.  The plot shows, for all particles ending up projected onto the central 2\,pc on the sky, the relative importance for changing the mass function of collisions between stars of masses $M_{\star,1}$ and $M_{\star,2}$.  The colour scale shows $\ud^2 N_{\rm coll} / \ud \ln M_{\star,1}\,\ud \ln M_{\star,2}$, where $N_{\rm coll}$ is the total number of collisions between stars in those two mass bins.  Each collision is shown twice on the figure.}
    \label{fig:collMap}
\end{figure}

\subsection{Effects on the mass function}

Figure~\ref{fig:collMap} shows the relative importance of collisions between pairs of main-sequence stars at different mass pairings.  For any mass of main-sequence star the most common main-sequence impactor is from the lowest-mass bin, simply because those stars are most numerous, but the mass which they bring with them is relatively small.  Plotting $\ud^2 N_{\rm coll} / \ud \ln M_{\star,1}\,\ud \ln M_{\star,2}$ shows which mass impactors have the most effect on a given species.  The plot shows that, for a given target star of mass $M_1$, the fractional change in mass as a result of collisions is roughly equal for all impactors with mass $M_2<M_1$.  The effect of impactors with $M_2>M_1$ is negligible. The peak effect is seen around 0.8 to $1\,\msun$, which is where the effects of collisions between main-sequence stars is having the most effect on the mass function: these are the highest-mass stars to be on the main sequence for the entire simulation duration.

\begin{figure}
    \includegraphics[width=\columnwidth]{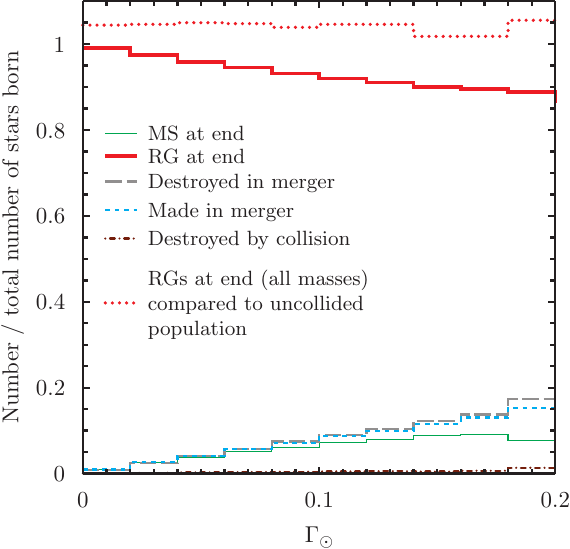}
    \caption{Stellar population fractions at the end of the simulation for stars of initial mass around the turnoff mass as a function of interaction parameter $\Gamma_\odot$.  Stars with initial masses between $0.9375\,\msun$ and $0.9875\,\msun$ are shown with solid and dashed lines.  The various lines show the number of main-sequence stars (thin solid green line), and red giants (thick solid red line), the number of stars destroyed by mergers (long-dashed grey line), the numbers made in mergers of lower-mass stars (blue short-dashed line) and the number destroyed by collisions with compact objects (brown dot-dashed line). Finally, the red dotted line shows the ratio between the total number of red giants in all mass bins at the end of the simulation and the total number in an uncollided population.
    }
    \label{fig:giantDestructionFraction}
\end{figure}

Figure~\ref{fig:giantDestructionFraction} shows the final state of stars that initially had masses between $0.9375\,\msun$ and $0.9875\,\msun$, binned by interaction parameter $\Gamma_\odot$.  For a non-interacting population all of these stars would be giants.  For an interacting superparticle, three factors affect the population.  First, collisions with other main-sequence stars remove stars from the mass bin (grey long-dashed line).  Secondly, the population is replenished at almost the same rate by the creation of stars in collisions.  Many of these stars, however, do not manage to evolve onto the giant branch by the end of the simulation and are still on the main sequence at the end (green thin solid line).  Finally, some of the initial population is destroyed by collisions with compact objects (brown dot-dashed line).  It can be seen that the rate of destruction by compact objects is small compared to the rate of stellar mergers, but that the population removed by mergers is largely replenished.  Indeed, if red giants at masses other than the bins around the turnoff mass are included (red dotted line) then the number of visible giants is actually slightly increased by collisions.

\begin{figure}
    \includegraphics[width=\columnwidth]{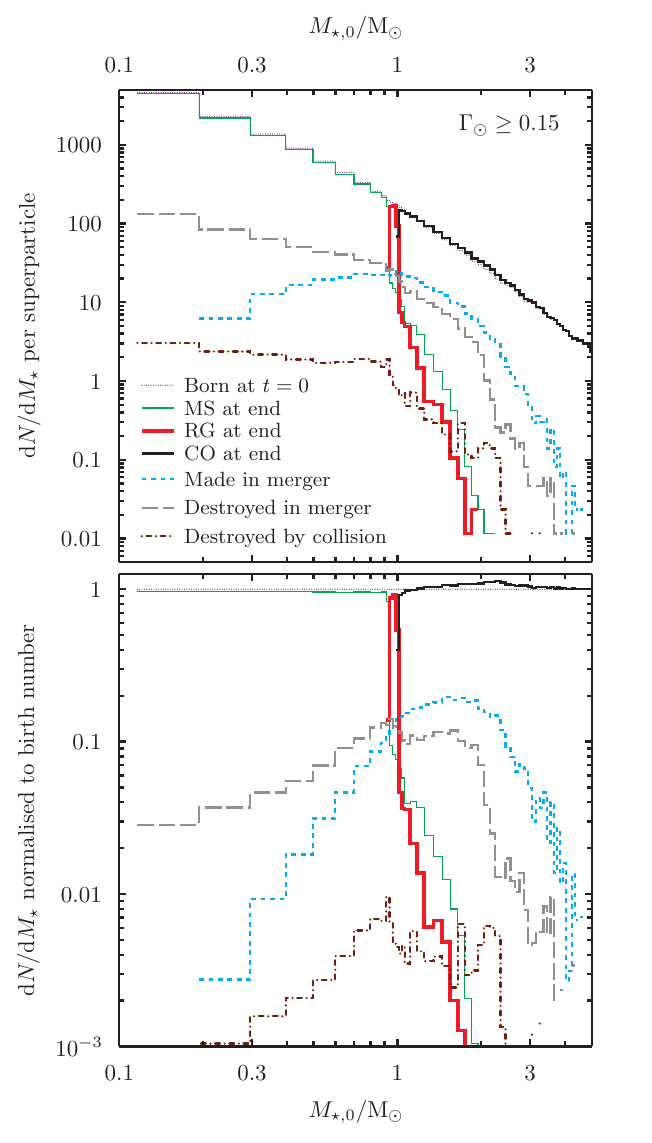}
    \caption{Effects of close encounters on the numbers of stars as a function of initial stellar mass, $M_{\star,0}$.  {\it Top:} Number of stars of different types per unit mass at the end of the simulation.  The thin, dotted purple line shows the initial population, the thin solid green line the stars that are on the main sequence at the end of the model, the thick solid red line red giants, and the black solid line compact objects (WDs, NSs and BHs).  The blue short-dashed line shows the stars produced by mergers, and the grey long-dashed line the stars consumed by mergers.  The brown dash-dotted line shows the stars destroyed by collision with a compact object.  {\it Bottom:} As top, but normalised by the initial number of stars of each mass.}
    \label{fig:birthsDeaths}
\end{figure}

Figure~\ref{fig:birthsDeaths} gives an alternative view of how mergers and collisions with compact objects sculpt the stellar population.  This figure shows the mass distributions of different types of stars and of the effects of stellar mergers and collisions at the end of the simulation. The thin dotted purple line shows the initial stellar population: for a non-interacting population this would be equal to the main sequence population (thin solid green line) below the turnoff, giants (thick solid red line) around the turnoff and compact objects (black solid line) above the turnoff.  The short-dashed blue line shows the stars created by mergers, and the long-dashed grey lines the stars thus destroyed.  The main effect of collisions is to convert low-mass stars into higher-mass stars, with the largest relative increase in population being around $2-3\,\msun$.  Almost all of these stars have evolved by the end of the simulation, however, which leads to an increase in the number of white dwarfs descended from such stars (the bulge in the black line).  At the present-day turnoff mass, more stars are destroyed than created by collisions, which leads to a significant reduction in the number of red giants around the turnoff; however, this is compensated by higher-mass red giants.  Finally, it can be seen that the effect of collisions between stars and compact objects is negligible for all stellar masses.

Figures~\ref{fig:giantDestructionFraction}~and~\ref{fig:birthsDeaths} show that, even for the most interactive superparticles, the effects on the visible population of stars are modest.  At no values of $\Gamma_\odot$ do we see more than a 10\% change to the mass function, and the total number of giants is barely changed by stellar encounters.  Part of the reason for this is because the surface escape speeds of the stars are similar to the relative velocities of the stars in our simulations.  This means that, with respect to the stellar velocity dispersion, the encounter rates are at a minimum.  Hence, despite the very high number densities of stars, direct encounters between single stars are relatively inefficient at modifying the stellar population.

\begin{figure}
\begin{center}
\includegraphics[width=.9\columnwidth]{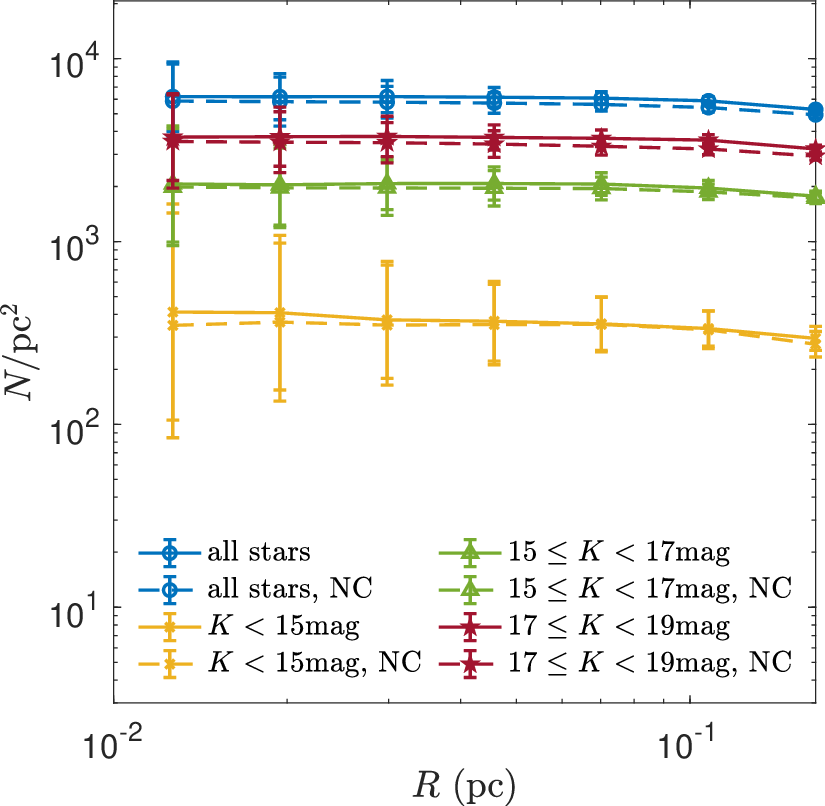}
\end{center}
\caption{Projected number density profile for the stars in 3 different magnitudes bins and for all entire stellar sample. Solid lines are for the NSC modified by the stellar collisions, while dashed lines are for the system not affected by collisions. The binning and radial extension are similar to those adopted in by \protect\citet{Schoedel20} for their figure 12.}
\label{fig:DensityMag}
\end{figure}

\subsection{Effect on the luminosity function}

Figure \ref{fig:DensityMag} shows the number density profiles for the stars in three different magnitude bins and for the entire stellar sample. The density and its errors are calculated by randomly sampling 1000 times the over-sampled final snapshot.

The collisions do not alter the stellar distribution within 0.2~pc (0.5'') from the central SMBH significantly. The stars in the brighter bin seem to move on a steeper profile, however the profiles obtained with and without collisions are compatible within the errors. Fainter, and probably older and more relaxed, stars in the Galactic NSC seem to show a steep profile (\citealt[see figure 12 in][]{Schoedel20} and figure 4 in \citealt{Habibi19}).
Given the slight collision-related changes in the different magnitude bins, we conclude that the luminosity function is not significantly affected by the collisions.

\section{Formation and evolution of fuzzballs}
\label{sect:fuzzballs}

One prediction of our results is that red giant stars in the Galactic Centre should undergo periodic collisions with objects that are much more compact than their envelopes: main-sequence stars, white dwarfs and black holes.  These collisions are typically neither destructive of the giant nor, at the relative velocities found in the Galactic NSC, sufficiently dissipative for the compact star to be captured by the giant.  Instead the giant suffers some mass loss, and some of that mass emerges bound to the compact star.  Here we discuss the structure and observability of compact stars with tenuous accreted envelopes, which we refer to as {\it fuzzballs}.

\subsection{Structure and evolution of fuzzballs}

We assume that the captured gas (which we refer to as the {\it envelope}) reaches hydrostatic equilibrium within a free-fall time-scale.  Thereafter its properties are governed by the equations of stellar structure.  We solve these equations to connect its observable surface properties -- the radius $R$, effective temperature $\Teff$ and luminosity $L$ -- with the mass of the compact star $M_{\rm c}$, the mass of the envelope $M_{\rm env}$ and the thermal time-scale of the fuzzball envelope, $\tau_{\rm th,env}$.  Our analysis is based on Section~7.3.4 of \citet{EldridgeToutBook}, with appropriate modifications.  We make the assumption that the envelope is unstable to convective motions, and hence isentropic with $\gamma=5/3$: this allows us to write the pressure $P$ in terms of the density $\rho$ as
\begin{equation}
P=K\rho^{5/3},
\label{eq:isentropy}
\end{equation}
where $K$ is a constant to be determined.  We further assume that $M_{\rm env} \ll \Mc$, as seen in SPH simulations of compact stars colliding with giant envelopes: hence we can take the enclosed mass $m=\Mc$ in the equation of hydrostatic equilibrium, which becomes
\begin{equation}
\frac{\ud P}{\ud r} = \frac{5}{3}K\rho^{2/3}\frac{\ud \rho}{\ud r} = -\frac{G\Mc\rho}{r^2},
\end{equation}
where $r$ is the radial co-ordinate, $G$ is the gravitational constant and we have substituted Eq.~\ref{eq:isentropy}.  Rearranging and integrating with respect to $r$ gives the density as
\begin{equation}
\rho(r) = \left\{\rho_{\rm s}^{2/3} + \frac{2G\Mc}{5KR}\left(\frac{R}{r}-1\right)\right\}^{3/2},
\end{equation}
where $\rho_s$ is the surface density.  This allows us to obtain the envelope mass as
\begin{equation}
M_{\rm env} = \int_{\Rc}^R 4\uppi r^2 \rho(r)\ud r
\label{eq:Menv}
\end{equation}
where $\Rc$ is the radius of the base of the envelope.  The value of $\Rc$ is largely immaterial because, close to the core, the enclosed mass scales as $r^{3/2}$: hence the inner envelope's mass is too small to have much effect on the fuzzball's bulk properties.

It remains to define boundary conditions.  In addition to the isentropic relation we apply an ideal gas equation of state at the surface which allows us to obtain an expression for $K$ in terms of the surface density $\rho_{\rm s}$:
\begin{equation}
K = \frac{\mathcal{R}\Teff}{\mu\rho_{\rm s}^{2/3}}
\end{equation}
where $\mathcal{R}=8.314\,{\rm kJ\,kg^{-1}\,K^{-1}}$ is the gas constant and the mean molecular weight $\mu\approx0.7$ for a typical population I star envelope.  Finally, we apply a grey surface boundary condition for the surface pressure $P_{\rm s}$ and opacity $\kappa$:
\begin{equation}
\frac{2}{3}\frac{GM}{R^2} = \kappa(\rho_{\rm s},\Teff)P_{\rm s} = \frac{\mathcal{R}\rho_{\rm s}\Teff\kappa(\rho_{\rm s},\Teff)}{\mu}.
\end{equation}
We first choose $R$ and $\Teff$, and then solve this boundary condition iteratively using a bicubic spline fit to the opacities of \citet{Ferguson+05} to obtain $\rho_{\rm s}$.  The envelope mass is then obtained by integrating Eq.~\ref{eq:Menv}.  Finally, we obtain the internal energy of the envelope as
\begin{equation}
U = \int_{\rm env}\frac{3}{2}\frac{\mathcal{R}T}{\mu}\ud m = 6\uppi K\int_{\Rc}^{R}r^2\rho(r)^{5/3}\ud r,
\end{equation}
where we have again used the ideal gas law, and the gravitational potential energy of the envelope by
\begin{equation}
\Phi = -\int_{\rm env}\frac{G\Mc}{r}\ud m = -4\uppi G\Mc\int_{\Rc}^{R}r\rho(r)\ud r.
\end{equation}
The thermal time-scale for the envelope is then given by
\begin{equation}
\tau_{\rm env,th} = \left|\frac{\Phi+U}{L}\right|
\end{equation}
where the luminosity $L$ is given by the usual surface boundary condition $L=4\uppi R^2\sigma_{\rm sb}\Teff^4$ and $\sigma_{\rm sb}$ is the Stefan-Boltzmann constant.  Our calculation of the time-scale implicitly neglects any luminosity provided by the central compact star or by nuclear burning at the base of the envelope; the sole source of energy is the contraction of the envelope.

\subsubsection{Results}

\begin{figure}
    \includegraphics[width=\columnwidth]{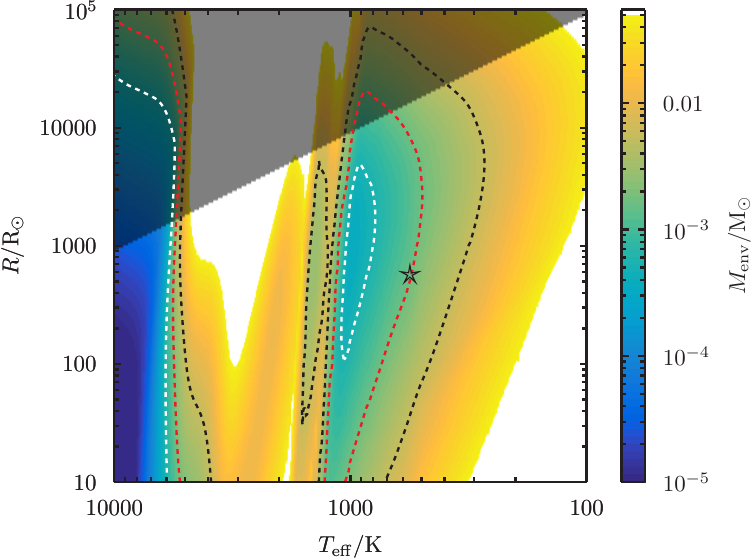}
    \caption{Envelope mass $M_{\rm env}$ as a function of surface temperature $T_{\rm eff}$ and radius $R$ for fuzzballs with a white dwarf core.  Dark regions of the plot have total envelope energies greater than zero; i.e. the envelope is not bound.  The white, red and black dashed contour lines show envelope mass ratios $M_{\rm env}/\Mc$ of $10^{-3}$, $3\times10^{-3}$ and $10^{-2}$ respectively.  White regions of the figure show envelope masses $M_{\rm env}>0.1\,\Mc$, where our model is no longer self-consistent.  The black star shows the location of G2 as measured by \citet{Witzel+14}.
    }
    \label{fig:WDfuzzball-dm}
\end{figure}

We start by considering the properties of fuzzballs with white dwarf cores, since these should be one of the more common cases.  Figure~\ref{fig:WDfuzzball-dm} shows the envelope mass $M_{\rm env}$ as a function of $T_{\rm eff}$ and $R$.  Black, red, and white dashed contour lines show envelope masses of $10^{-3}$, $3\times10^{-3}$ and $10^{-2}$ times the white dwarf mass; these represent the upper end of the distribution of envelope masses found in collisions in our models.  A given fuzzball has a fixed envelope mass and hence will follow a contour line, radiating away its internal energy as it shrinks.  The structure in the diagram is largely set by the complex dependence of the surface opacity on temperature and density.  It shows three branches in the figure -- the branch at around 1500\,K corresponds very roughly to the classical Hayashi track.  Our somewhat cavalier treatment of the inner boundary condition frees us from the Hayashi limit, however, and permits an additional, cooler, set of solutions.  These intersect nicely with the properties of the G2 cloud inferred by \citet{Witzel+14} ($T_{\rm eff,G2}=560\,{\rm K}$ and $R_{\rm G2}=570\,\Ro$). We infer that if G2 is a WD-cored fuzzball it must have an envelope mass of $1.5\times10^{-3}\,\Mo$.

\begin{figure}
    \includegraphics[width=\columnwidth]{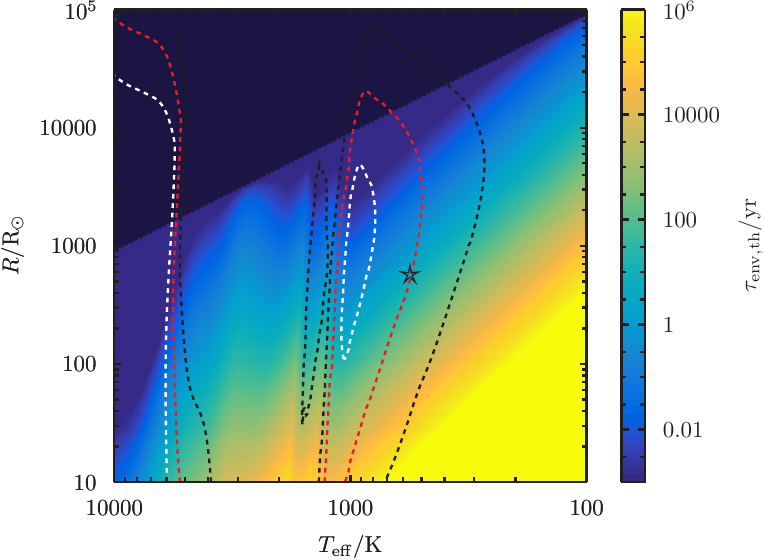}
    \caption{Thermal time-scale $\tau_{\rm env,th}$ as a function of $T_{\rm eff}$ and $R$ for fuzzballs with a white dwarf core. Black regions of the plot have total envelope energies greater than zero; i.e. the envelope is not bound.  The white, red and black dashed contour lines show envelope mass ratios $M_{\rm env}/\Mc$ of $10^{-3}$, $3\times10^{-3}$ and $10^{-2}$ respectively.  The black star shows the location of G2 as measured by \citet{Witzel+14}.
    }
    \label{fig:WDfuzzball-tau}
\end{figure}

While the surface properties of our model of G2 as a WD-core fuzzball are compatible with the observations, the model is excluded by the inferred thermal time-scale.  Figure~\ref{fig:WDfuzzball-tau} shows the thermal time-scale as a function of $T_{\rm eff}$ and $R$, with the star again marking the location of G2; our model at this point in the diagram has a thermal time-scale of 4.4\,yr.  This implies that G2 would have shown significant evolution of its surface properties over the last few years, which has not been observed, and that it was still in close proximity to the site of the close encounter.  In addition, the total mass is probably insufficient for G2 to have survived tidal disruption during its close passage by Sgr~A*.

\begin{figure}
    \includegraphics[width=\columnwidth]{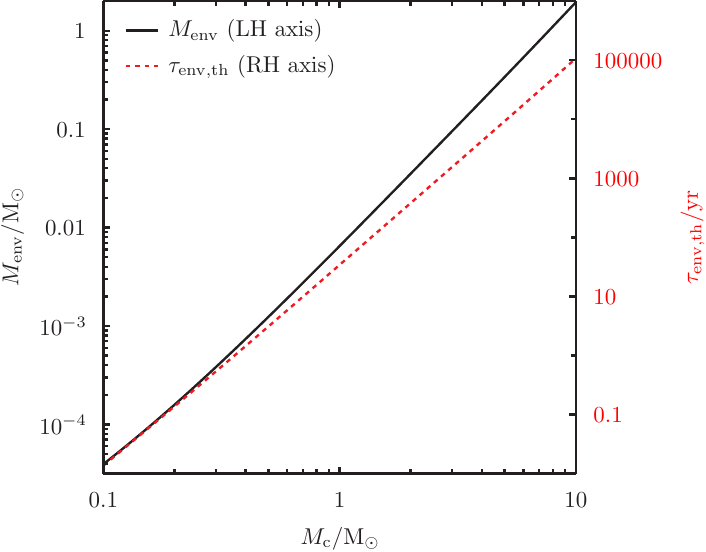}
    \caption{Properties of fuzzballs that have the same temperature and radius as G2.  The envelope mass $M_{\rm env}$ (black solid line, left-hand ordinate axis) and thermal time-scale $\tau_{\rm env,th}$ (red dashed line, right-hand ordinate axis) are plotted as functions of the compact star mass $\Mc$.  
    }
    \label{fig:fuzzball-properties}
\end{figure}

Figure~\ref{fig:fuzzball-properties} shows how the envelope mass and thermal time-scale of fuzzball models for G2 vary as a function of $\Mc$.  As can be seen, models start to become compatible with observational constraints on the thermal time-scale from $\Mc\approx 2\,\Mo$.  This rules out WDs and probably all but the most massive NSs as cores of a putative G2 fuzzball.  Black holes are still compatible, as are intermediate-mass main-sequence-star cores, but in the latter case more careful modelling would be required to take into account the effect of the star's own luminosity.

\begin{figure}
    \includegraphics[width=\columnwidth]{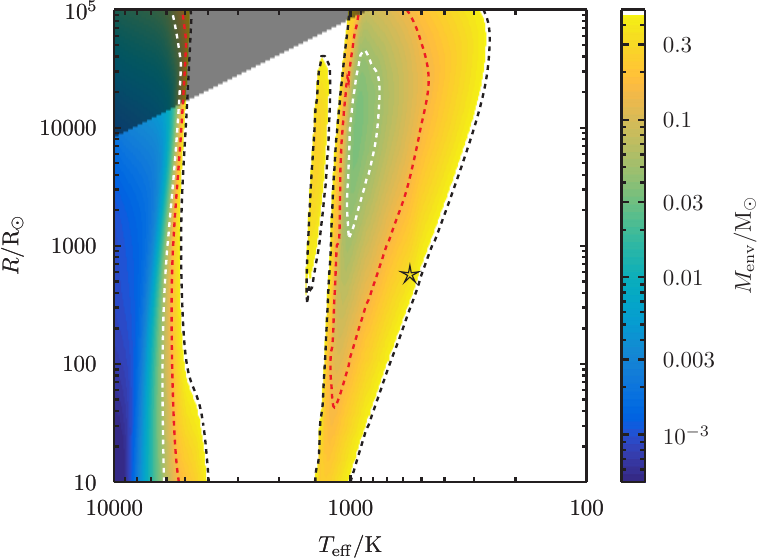}
    \caption{As Figure~\ref{fig:WDfuzzball-dm}, but for fuzzballs with $5\,\Mo$ black hole cores.  White, red and black dashed contour lines show envelope mass ratios $M_{\rm env}/\Mc$ of 0.01, 0.3 and 0.1.}
    \label{fig:BHfuzzball-dm}
\end{figure}

\begin{figure}
    \includegraphics[width=\columnwidth]{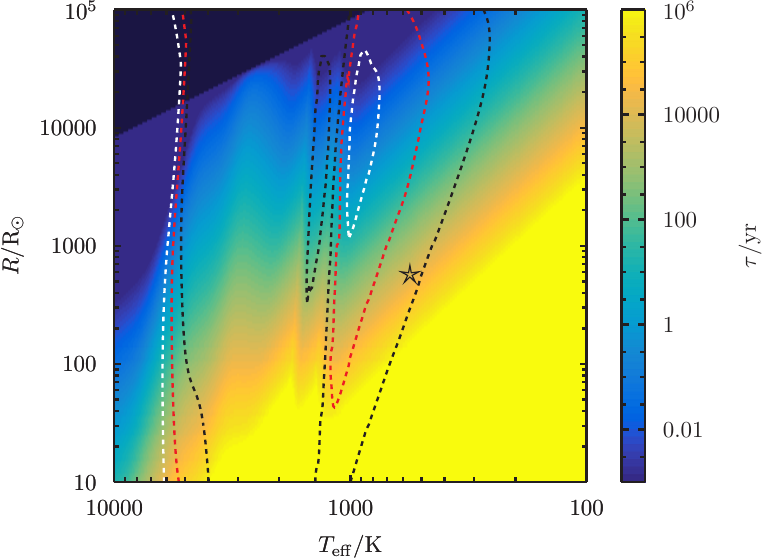}
    \caption{As Figure~\ref{fig:WDfuzzball-tau}, but for fuzzballs with $5\,\Mo$ black hole cores.  White, red and black dashed contour lines show envelope mass ratios $M_{\rm env}/\Mc$ of 0.01, 0.3 and 0.1.}
    \label{fig:BHfuzzball-tau}
\end{figure}

For completeness we include plots of the envelope mass (Figure~\ref{fig:BHfuzzball-dm}) and thermal time-scale (Figure~\ref{fig:BHfuzzball-tau}) for fuzzballs with central $5\,\Mo$ black holes.  In this case a larger ratio of envelope to core mass is required to match the G2 surface boundary conditions.  As a result of the larger envelope masses the thermal time-scales are greatly increased.  A BH fuzzball with the surface properties of G2 requires an envelope of $0.34\,\Mo$ and has a thermal time-scale of 9400\,yr, compatible with the properties of G2.

\subsection{Fuzzball formation rate}

\begin{figure}
\begin{center}
\includegraphics[width=.9\columnwidth]{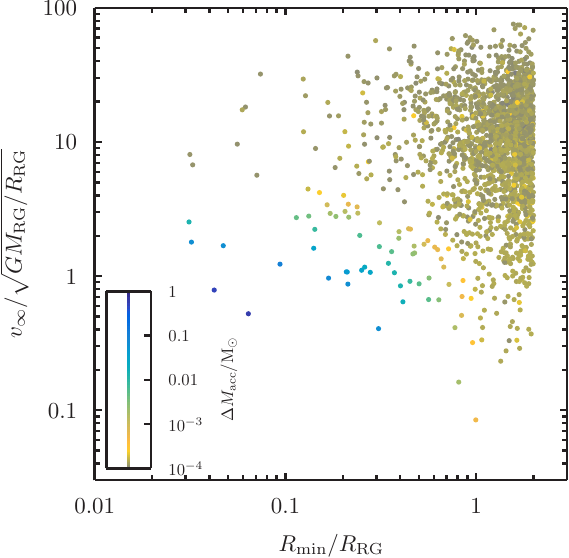}
\end{center}
\caption{Mass accreted by white dwarfs in collisions and close encounters with giant stars.  Points are plotted against the ratio of the closest approach radius to the giant surface radius, $R_{\rm min}/R_{\rm RG}$ (abcissa) and the ratio of velocity at infinity to stellar surface velocity, $v_{\infty}/\sqrt{GM_{\rm RG}/R_{\rm RG}}$.  Only close encounters at relatively low velocities accrete significant quantities of mass.  All encounters in the central projected 2\,pc and last Gyr of the simulation are shown.}
\label{fig:WDfuzzball-macc}
\end{figure}

Encounters between giant and compact stars are fairly common in our model.  We look at the average number of encounters in the whole central projected 2\,pc during the last Gyr of the simulation, after the NSC is fully assembled.  We find a total of 2100 WD--RG encounters, 65 NS--RG encounters and 15 BH--RG encounters.  Encounters with low-mass main-sequence stars are even more common, mostly with MS stars in the lowest mass bin.  However, most of the encounters lead to very little mass accretion.  Figure~\ref{fig:WDfuzzball-macc} shows the mass accreted by white dwarfs in encounters with giants in the final Gyr of our model.  The vast majority of encounters are fast and wide, since the the target presented by a giant is larger at larger radii, and the velocity dispersion in the central cluster is significantly larger than the surface escape speeds of the giants.  Only the rare, slow, close encounters accrete sufficient mass to form G2-like clouds, with a total rate of about 60 per Gyr.  Given the very short lifetimes of WD fuzzballs it is clear that G2 cannot be explained in this way.

\subsubsection{Frequency of harassment of giants}
\begin{figure}
\includegraphics[width=\columnwidth]{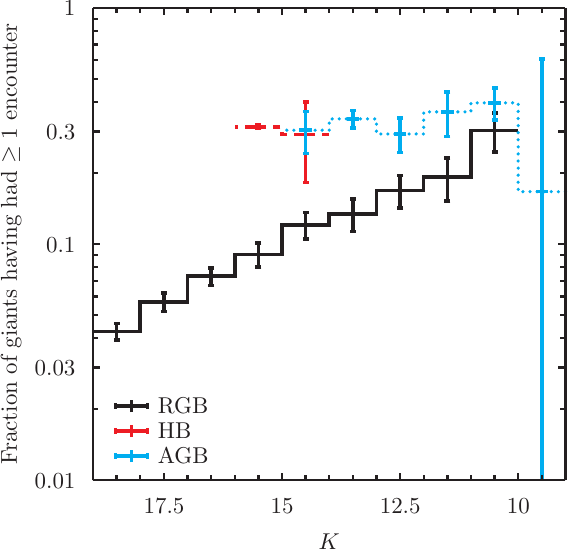}
\caption{Fraction of giants visible in central projected 2pc at the end of our simulation that have had at least one encounter with a compact star.  The fraction is plotted as a function of apparent $K$-band magnitude.  The black solid line shows stars on the first giant branch (RGB), the red dashed line stars in the red clump (HB) and the blue dotted line stars on the asymptotic giant branch (AGB).  Error bars show the $\sqrt{N}$ uncertainties from the number counts in each bin.}
\label{fig:giantEncounterFrac}
\end{figure}

A corollary to our previous statement is that the fraction of giant stars that have had an encounter with a compact star since leaving the main sequence is potentially rather large.  Figure~\ref{fig:giantEncounterFrac} shows the fraction of giant stars that have had at least one encounter with a compact star, split into bins by apparent $K$-band magnitude.  The figure shows that, as stars evolve up the first giant branch, the collision fraction increases steadily up to around one quarter of giants by the top of the RGB.  By the tip of the AGB about 40\% of giants will have had an encounter.  It is beyond the scope of this paper to study in detail the effect of these encounters on the giants, but  they will include a reduction in the giant lifetime, and a temporary increase in luminosity from deposition of thermal energy by the passing stars.

% {\blue
% Note: For fuzzball evolution, considering a uniform cube of gas of length $L$, $L=\sqrt{\kappa M/\tau}\approx 100\,{\rm AU}$ if $\kappa=1\,{\rm g\,cm^{-3}}$, $\tau=1$, $M=10^{-3}\,\msun$.
% }

\label{sect:discussion}
NSCs are among the densest stellar systems in the Universe and, as such, their stars should undergo frequent interactions.
Stellar collisions have been addressed in a number of studies, relying on analytical and observational based models of the Galactic nucleus \citep{Davies98, Alexander99, BaileyDavies99, Dale+09}.
In this work we tackle this problem using the phase-space distribution of stars in a Milky Way-like NSC, built through the infall and merger of GC-like massive clusters \cite[see e.g.][]{PMB14,Tsatsi17, Abbate18}. The NSC formed in this $N$-body simulations show morphological (e.g. flattening) and kinematic (e.g. rotation, velocity dispersion) properties comparable to those of the Galactic NSC.
After converting each $N$-body particle into a stellar population, we applied a stellar evolution routine and collision prescriptions derived from the literature to study the cumulative effects of the stellar encounter history at the Galactic Centre. We took into consideration collisions of main sequence and red giant stars with compact objects (WDs, BHs and NSs), as well as collisions of main sequence stars with other main sequence stars and red giants, while we neglected rare events such as collisions between red giant stars and between pairs of compact objects.

The most common encounters in our models are between pairs of main-sequence stars. Stars of a given mass are only significantly affected by collisions with less massive impactors, and the peak effect on the mass function is observed for stars with masses between 0.8 and 1\,$M_\odot$, i.e. for the highest-mass main sequence stars whose lifetime is not less than the simulation length.   However, around this mass the mergers remove and create stars at almost the same rate, so at the present day the red giants whose progenitors were destroyed by collisions are replaced by higher-mass red giants whose progenitors were born in mergers.

Furthermore, we find that the rate of collisions between stars and compact objects is so low that their effect on the mass function is negligible. The mass function of the visible stars in the Galactic NSC should, therefore, only be slightly affected by stellar encounters: we see changes smaller than 10\%.  The final core-like density profile of our NSC is not significantly altered by the collisions. Stellar encounters marginally affect the more luminous stars but leave the luminosity function of the stars in the NSC mostly unchanged.

Our NSC formation scenario leads to a central core at the end of the $N$-body simulation. We, therefore, do not probe the role that collisions might play in disrupting a cusp \citep{Davies98, Alexander99, BaileyDavies99, Dale+09, AS14, AS20}. Analysis of a static cuspy NSC model with properties similar to those of the Galactic nucleus suggest that, in this case, stellar encounters could have a non negligible impact on the luminosity function and spatial distribution of the observable stars. However, a more sophisticated analysis of a time-evolved MW-like NSC showing a central stellar cusp \citep{GaSc18, Schoedel20, Habibi19} will be necessary to test this scenario in detail.

We follow the evolution of fuzzballs -- compact stars that have accreted a tenuous envelope in collisions with red giants -- and conclude that the properties of the enigmatic G2 object observed at the Galactic centre could be explained as the result of an encounter between a red giant and a black hole. On the other hand, we exclude the possibility that G2 could be a WD fuzzball, given the short lifetime of these objects.  
Our collision rates make the BH-fuzzball model for G2 very unlikely, as our model has a core number density of black holes in the central $0.5\,{\rm pc}$ of only ${\rm 1400\,BHs\,pc^{-3}}$. The collision rate would be much higher if a cusp of black holes is present in the central NSC \citep[see e.g.][for theoretical and observational studies on the presence of massive stars and SBHs at the Galactic centre]{Miralda00, Freitag06, alexander09, Hailey18, Generozov18, Davies20}. The number of black holes in the galactic centre could also be enhanced by local star formation. The observed population of massive young stars seen today will likely produce up to $100$ black holes. If this mode of star formation occurs every $100$~Myr or so \citep{AS14}, then in total, this mode of star formation will add roughly $10^4$ black holes in the central regions thus enhancing the population by a factor of roughly ten. In turn if some of these black holes are more massive (around $20\,\msun$), then the thermal relaxation times for any fuzzball produced will be commensurately longer (as shown in Figure \ref{fig:WDfuzzball-dm}). Thus with a population of black holes enhanced in this way, the chance of seeing G2 today could be a few per cent or higher.

Another possibility is fuzzballs produced by collisions of giants with intermediate-mass stars.  In this case the luminosity provided by the central star might be expected to compensate for the loss of energy from the photosphere of the tenuous envelope, since G2's luminosity is roughly the same of a $2\,\msun$ main-sequence star.  However, it is by no means clear that the structure of a central core and tenuous envelope would persist over multiple thermal timescales, and modelling the evolution of such objects would require a solution of the full set of equations of stellar evolution which is beyond the scope of this paper.

Another possible route to form G2 is collisions between main-sequence stars.  Collisions deposit large quantities of kinetic energy which produce a distended star, possibly with surface properties similar to G2.  We see seven MS--MS collisions per Myr in the final $10^9\,{\rm yr}$ of simulations, which suggests that this is an interesting avenue to explore in further work.   Another route that may lead to repeated grazing MS--MS collisions is collisions between successive EMRIs, one of them spiraling in towards the black hole as it emits gravitational waves, while the other spirals outwards as it transfers mass stably to the black hole \citep{MetzgerStone17}; however, such collisions, taking place at a significant fraction of the speed of light, would not produce G2-like objects.

In our model, about 40\% of the giants undergo at least one encounter by the end of their AGB phase. Such encounters may shorten their lifetimes, and temporarily increase their luminosity due to the energy deposited by the interacting star. Further SPH simulations and longer-term follow-up with stellar evolution calculations will be necessary to follow the details of such events, as well as to to better understand the formation and evolution of fuzzballs.

Our results might suggest that stellar encounters are unlikely to be of any significance to the stellar population at the Galactic Centre, and that further work on this complex edifice of encounter simulations promises to be of little interest.  However, as aluded to above, various assumptions that we have made conspire to make this a very pessimistic scenario.  The core-like structure of our NSC reduces the central density of the system and the presence of the supermassive black hole throughout the evolution in effect enforces the high velocity dispersion that reduces the stellar encounter rate.  Secondly, we inject the stars into the simulation when the cluster enters the field, but ignore their previous evolution.  Globular clusters undergo dynamical encounters which not only modify their stellar populations by collisions and coagulations, but also cause segregation of the more massive stars into the centres of the clusters.  Failing to include these effects means that our results are inherently a pessimistic estimate of the effects on the stellar population from encounters.  

We plan to refine our models to investigate the likely effects of some of these assumptions in a subsequent paper. In particular, we will combine our NSC formation simulations with realistic GC models \citep[e.g. obtained with MOCCA][]{Giersz08,Giersz13} to provide more accurate predictions on the characteristics of the stellar populations in the Galactic nucleus. We will also explore the collisional evolution of a cusp-like NSC with the same properties of the Galactic NSC to once and for all check if the collisions can flatten the density profile.

\section{Conclusions}
\label{sect:conclusions}
In this work we attempt to quantify the effect that close encounters between stars has had on the stellar populations in the Milky Way nuclear star cluster (NSC).  NSCs should witness significant rates of stellar encounters owing to their high number densities.  We model the stellar evolution and collision history of a Milky Way-like NSC built through the infall of stellar systems similar to GCs, basing our simulations on $N$-body models of NSC formation through the merger of multiple massive stellar clusters.  We find that the most common encounters to be expected in the NSC are collisions between pairs of main-sequence stars.  

Stellar collisions and mergers have little effect on the observable stellar populations, both in terms of their mass and luminosity functions.  There are three reasons for this.  First, our $N$-body models form a core rather than a cusp, which limits the central number density that we reach and hence the collision rate.  Second, the presence of a supermassive black hole from the beginning of the simulation leads to a high velocity dispersion, similar to the surface escape speeds of the interacting stars.  This coincidence minimises the rate of stellar encounters.  Third, while collisions between main-sequence stars can deplete the population of bright giants by accelerating the evolution of their progenitors, it can also enhance the population by accelerating the evolution of lower-mass main-sequence stars.  For modest collision rates these effects cancel and the population of visible giants is almost unaffected.

Additionally we investigate the possibility that a main-sequence star or compact remnant could  accrete a tenuous envelope when passing through the outer layers of a red giant.  We test whether such objects, which we refer to as {\it fuzzballs}, could explain G2, a cloud observed to have made a close passage past the MW SMBH and whose nature is still debated.  We find that fuzzballs where the accretor is a white dwarf or low-mass main-sequence star are ruled out on the grounds of their thermal time-scales being too short.  A black-hole fuzzball has a compatible thermal timescale but in our model the formation rates of such objects are too low to convincingly explain G2.

We plan to overcome the limitations of our $N$-body simulations and analysis by considering more realistic GC models obtained e.g. in Monte Carlo simulations \citep[e.g. MOCCA][]{Giersz08, Giersz13}. We will also consider models in which the underlying density profile of the NSC is steeper, which could lead towards a higher rate of collisions and to a more significant effect on the stellar populations at the Galactic centre. 

\section*{Acknowledgements}
The authors would like to thank Jim Dale for sharing his code upon which we built our coagulation calculator.  We would also like to thank Abbas Askar, Christopher Tout and Anja Feldmeier-Krause for helpful discussions.  
AMB is supported by the Swedish Research Council (grant 2017-04217).

Calculations presented in this paper were enabled by resources provided by the Swedish National Infrastructure for Computing (SNIC) at LUNARC, partially funded by the Swedish Research Council through grant agreement no. 2018-05973 and partially funded by grants from the Royal Fysiographic Society of Lund.

\section*{Data Availability}

Access to the $N$-body simulations on which this work is based \citep{Tsatsi17} and to the encounter histories that we present are possible by e-mail request to the corresponding author.  Results of stellar encounter simulations were drawn from the published literature \citep{Davies98, Alexander99, BaileyDavies99, Dale+09}.
 
% The inclusion of a Data Availability Statement is a requirement for articles published in MNRAS. Data Availability Statements provide a standardised format for readers to understand the availability of data underlying the research results described in the article. The statement may refer to original data generated in the course of the study or to third-party data analysed in the article. The statement should describe and provide means of access, where possible, by linking to the data or providing the required accession numbers for the relevant databases or DOIs.

\bibliographystyle{mnras}
\bibliography{paperI}

% Don't change these lines
\bsp	% typesetting comment
\label{lastpage}
\end{document}